\documentclass{aastex631}

\newcommand{\vect}[1]{\boldsymbol{#1}}
\newcommand\textred[1]{{\color{black}#1}}
\newcommand\textblue[1]{{\color{black}#1}}
\usepackage{amsmath}
\usepackage{algorithm}
\usepackage{algorithmic}
\usepackage{bbding}
\usepackage{booktabs}
\usepackage{hyperref}
\hypersetup{
    colorlinks=true,
    linkcolor=blue,
    filecolor=magenta,      
    urlcolor=cyan,
    }

\usepackage{xspace}
\newcommand\ratio{\texttt{ratio}\xspace}

\newcommand{\Acc}{\mathrm{Acc}}

\newcommand\cS{{\mathcal S}}
\newcommand\cC{{\mathcal C}}
\newcommand\cG{{\mathcal G}}
\newcommand\cP{{\mathcal P}}

\begin{document}

\title{Polarization Holes as an Indicator of Magnetic Field-Angular Momentum Alignment \textred{I. Initial Tests}}

\correspondingauthor{Xiaodan Fan and Hua-bai Li}
\email{xfan@cuhk.edu.hk and hbli@phy.cuhk.edu.hk}

\author{Lijun Wang}
\affiliation{Department of Statistics, The Chinese University of Hong Kong, Shatin, NT, Hong Kong SAR, China}
\affiliation{Department of Biostatistics, Yale University, New Haven, Connecticut, USA}

\author{Zhuo Cao}
\affiliation{Department of Physics, The Chinese University of Hong Kong, Shatin, NT, Hong Kong SAR, China}

\author[0000-0002-2744-9030]{Xiaodan Fan}
\affiliation{Department of Statistics, The Chinese University of Hong Kong, Shatin, NT, Hong Kong SAR, China}

\author[0000-0003-2641-9240]{Hua-bai Li}
\affiliation{Department of Physics, The Chinese University of Hong Kong, Shatin, NT, Hong Kong SAR, China}









\begin{abstract}

The formation of protostellar disks is still a mystery, largely due to the difficulties in observations that can constrain theories. For example, the 3D alignment between the rotation of the disk and the magnetic fields (B-fields) in the formation environment is critical in some models, but so far impossible to observe. Here, we study the possibility of probing the alignment between B-field and disk rotation using ``polarization holes'' (PHs). PHs are widely observed and are caused by unresolved B-field structures. With ideal magnetohydrodynamic (MHD) simulations, we demonstrate that different initial alignments between B-field and angular momentum (AM) can result in B-field structures that are distinct enough to produce distinguishable PHs. Thus PHs can potentially serve as probes for alignments between B-field and AM in disk formation.

\end{abstract}



\section{Introduction} \label{sec:intro}

A protostellar disk is a critical step in star formation and a major mystery. It is difficult to understand disk formation because turbulence, the source of disk AM, is not much more energetic than B-fields, so the B-fields can consume significant rotational energy, resulting in the so-called ``magnetic braking catastrophe'' (MBC, e.g., \citealt{Price2007, Hennebelle2008}) for disk formation. A number of approaches have been proposed to resolve MBC, including efficient ambipolar diffusion (e.g., \citealt{Masson2016, Tang_2018, galaxies9020041}), reconnection diffusion (e.g., \citealt{2012SSRv..173..557L}), \textred{Hall effect (e.g. \citealt{wursterCanNonidealMagnetohydrodynamics2016,tsukamotoBIMODALITYCIRCUMSLAR2015})} and field-rotation misalignment (e.g., \citealt{Li_2013}). In this work, we propose a method for probing the 3D (mis)alignment between B-fields and AMs, which will be referred as the ``BAM alignment''.

Even directly observing sky-projected 2D alignment between a disk and the vicinity B-fields is almost impossible currently, not to mention 3D BAM alignment. 2D B-field morphology observations depend on the polarization of thermal dust emission. However, neither interferometers nor single-dish telescopes can observe field morphology in a disk's vicinity. Interferometers filter out most of the disk envelopes, while single-dishes cannot resolve them. Interferometers may resolve the B-field within a disk after our knowledge of grain alignment and grain size population within disks is improved; but, even then, the B-field within disks should be just toroidal and tell nothing about BAM alignment prior to disk formation.

\begin{figure}
    \centering
    \includegraphics[width=\textwidth]{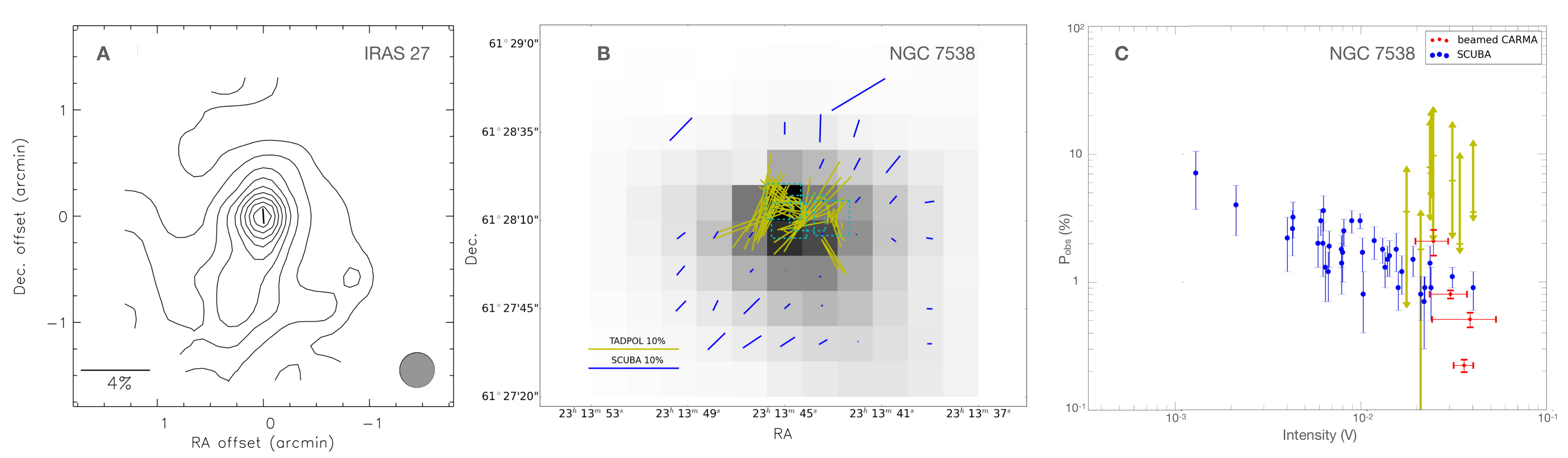}
    \caption{\textred{Examples of polarization holes. The thermal dust emission intensity is shown as contours in panel A, grayscale in B, and x-axis values in C. Polarizations are represented as line segments in A and B, and as y-axis values in C. Panel A and the blue data in B/C are from single-dish polarimetry observations. Most single-dish sub-mm polarimetry observations resemble the blue data in B/C, exhibiting an inverse correlation between polarization fraction ($P_{obs}$) and intensity known as a polarization hole. The condition of A — increasing $P_{obs}$ with intensity — is very rare. For IRAS 27, only the polarization at the intensity peak is high enough for being detected. The yellow data in B is from the interferometer CARMA, which has a much higher resolution than the blue data taken by JCMT/SCUBA, the resolution of which is indicated by the grey-scale pixel size in B. The blue data in C is also from JCMT, but the yellow data shows CARMA $P_{obs}$ versus the intensity of JCMT pixels that overlap with the CARMA detections for the purpose of comparing single-dish/interferometer $P_{obs}$ from the same lines of sight. Accordingly, a yellow error bar indicates the range of CARMA $P_{obs}$ covered by the same JCMT pixel. The blue error bars represent the measurement uncertainties of JCMT $P_{obs}$. Four synthesized JCMT beams (cyan dashed squares) in B are positioned to cover the majority of CARMA detections. The average JCMT intensity and the Stokes mean of CARMA $P_{obs}$ within the synthesized beam are represented as red data points in C.}}
    \label{fig:ph}
\end{figure}

Here we study the possibility of probing BAM alignment with so-called ``polarization holes'' (PHs). PHs are  widely observed negative correlations between cloud column density and the polarization fraction of thermal dust emission. An example of PH can be found from Figure 1. A large body of literature attributes PHs to the decreasing grain alignment efficiency (GAE) with increasing density. However, recently, we found that Alfv\'en Mach number, in fact, increases with density and turbulence becomes slightly super-Alfv\'enic in cloud cores (\citealt{Zhang_2019, Cao_2023}; Yuan \& Li in prep.). This would naturally correlate the complexity of B-field structure with density and explain PHs without the help from varying grain alignment efficiency (\citealt{alma991039385397203407}; Chan \& Li in prep.). Moreover, the GAE interpretation of PHs \textred{failed to explain the higher polarization fractions observed by interferometers when pointing at the ``bottom'' of a PH (Figure \ref{fig:ph}; \citealt{alma991039385397203407}).  If GAE decreased with increasing density, the interferometer-detected polarization fractions should be even lower. This is because interferometers not only better focus on the density peak with smaller beam size but also tend to filter out the larger-scale fore/background with lower densities. In fact, when \cite{alma991039385397203407} convolved interferometer detections with the single-dish beam size, he ``recovered''  the low single-dish polarization fractions (for example, see Figure \ref{fig:ph}C, red color). This directly illustrates that the hole is due to higher B-field complexity.}

As our first test of using PHs to access BAM alignment, we simulate ideal MHD cloud cores to study whether parallel and perpendicular BAM alignments result in distinct PH patterns. If this is the case, these simulated PH patterns can be utilized as a benchmark for evaluating the BAM alignment of observed PHs.  Due to the fact that core turbulence being slightly super-Alfv\'enic and being the source of disk AM, in the same way we set the rotational to B-field energy ratio. A detailed description of the simulation setup can be found in Section~\ref{sec:mhdsim}. Assuming a constant grain alignment efficiency, we can obtain PH patterns due to rotation-induced B-field structures. Multiple PH patterns are created for each BAM alignment by projecting along various lines of sight (LOS). By using statistical methods, we study whether the PH patterns from the same BAM alignment can be classified. The statistical method is introduced in Section~\ref{sec:method}, followed by the results in Section~\ref{sec:res} and discussion in Section~\ref{sec:discussion}. 

\section{MHD Simulations}\label{sec:mhdsim}
\subsection{Simulation Setup}
\begin{figure}[ht]
  \centering
  \includegraphics[width=0.5\textwidth]{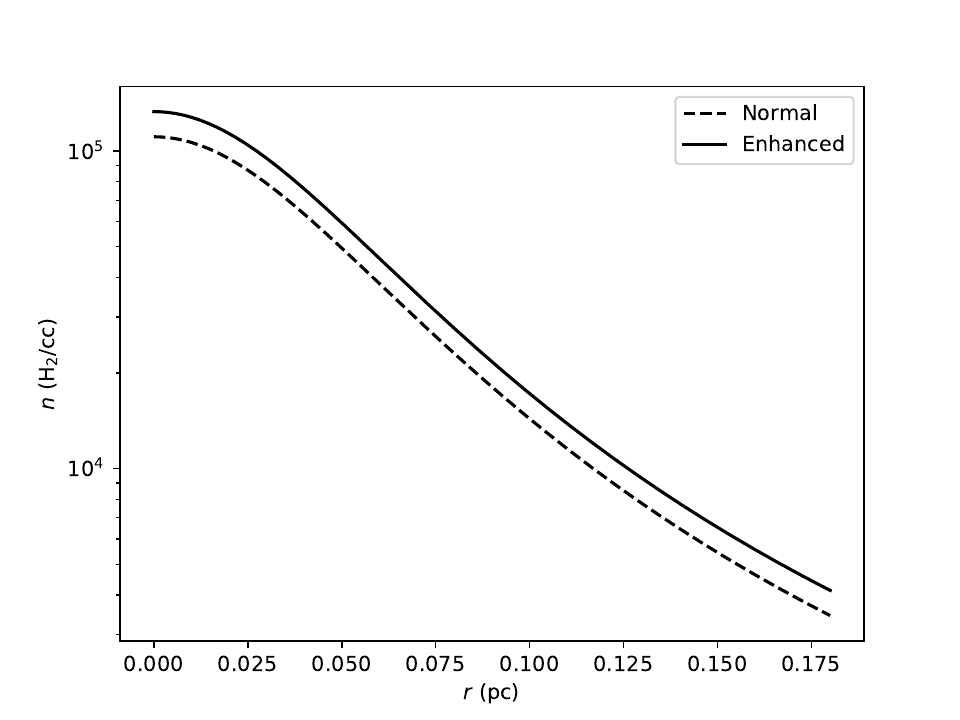}
  \caption{The normal and the 1.2 times enhanced density profile of Bonnor-Ebert Sphere (\citealt{10.1093/mnras/116.3.351,1955ZA.....37..217E}).}
  \label{fig:be_den_profile}
\end{figure}

Cloud cores have two critical parameters that we try to replicate in our simulations: a slightly supercritical gravitational energy \citep{2013MNRAS.436.3707L, 2021ApJ...917...35M} and a slightly super-Alfv\'enic kinetic energy (\citealt{Zhang_2019, Cao_2023}; Yuan \& Li in prep.). The former was simulated using a slightly enhanced Bonnor-Ebert sphere profile ($\rho_{BE}$, \citealt{10.1093/mnras/116.3.351,1955ZA.....37..217E}):
\begin{equation}
    \rho(r) = 1.2\ \rho_{BE}(r).
    \label{eq: BES_profile}
\end{equation}
The normal ($\rho_{BE}(r)$) and enhanced ($\rho(r)$) Bonnor-Ebert profiles are shown in Figure \ref{fig:be_den_profile}. The initial central density $\rho_c \sim 1.1 \times 10^5 \ \textrm{H}_2/\textrm{cc}$ and sound speed $c_s$ = 0.375 km/s are used to determine the profile $\rho_{BE}$.
A uniform $\sim$ 19.4 $\mu$G B-field is initialized in the z-direction. This gave a magnetic criticality of 3.5 within $r_{core}$ = 0.15 pc, which is a typical core size observed by $\sim$ 10-meter single dishes.

For kinetic energy, we simplify the rotation as solid-body with a radius $r_{rot}$ = 0.03 pc (see discussion in Section~\ref{sec:discussion}). In this regard, the simplification is valid since we do not intend to resolve B-field within the rotator. Instead, we are trying to determine how the rotator will affect the envelope field morphology, leading to PH patterns. Within $r_{rot}$, the rotation to magnetic energy ratio is set to 3. The rotation axis is set in the z-direction for the case of rotation parallel with the B-field and is in the x-direction for the perpendicular case.

The periodic simulation domain cube has a size of 0.36$^3$ pc$^3$ resolved by 256$^3$ cells. The Scorpio code \citep{chengScorpio2023} is adopted to evolve the system governed by the following ideal MHD equations:

\begin{equation}
  \begin{aligned}
    \frac{\partial \rho}{\partial t} + \nabla \cdot (\rho \vect{v}) &= 0 \\
    \rho(\frac{\partial}{\partial t} + \vect{v}\cdot \nabla)\vect{v}       &= \vect{J}\times \vect{B} - \nabla p - \nabla \phi \\
    \frac{\partial \vect{B}}{\partial t}                            &= \nabla \times (\vect{v} \times \vect{B})  \\
    p                                                        &= \rho c_s^2 \\
    \vect{J}                                                        &= \nabla \times \vect{B} \\
    \nabla \cdot \vect{B}                                           &= 0 \\
    \nabla^2\phi                                             &= 4\pi G \rho
  \end{aligned}\,,
  \label{eq: phole_mhd_equations}
\end{equation}
where $\rho$ and $p$ are mass density and thermal pressure, respectively; $\vect{v}$ and $\vect{B}$ are velocity and magnetic field vector, respectively; the sound speed $c_s$ is 0.375 km/s assuming an isothermal condition. $G$ is the gravitational constant. The simulations were continued for $\sim$ 0.34 Myr, till the density peaks started to violate the Truelove criterion \citep{Truelove_1997}.

\subsection{Simulation Result}\label{sec:sim_res}
\subsubsection{Sample Projections}\label{sec:sample_proj}

A quick-look of the results is shown in Figures \ref{fig:p1} and \ref{fig:p2}, where the same five projection directions are applied to the parallel and perpendicular BAM alignment cases, respectively. These five directions consist of two rotation axes (i.e., the x- and z-axes) and three evenly distributed directions between them (i.e., 22.5 degrees between adjacent directions). Well-developed rotational systems are nearly axially symmetric. Hence, sampling projection directions in the x-z plane is very close to sampling the entire space.

We label the projections by three parameters $(\ell, \theta, \phi)$. $\ell = 0^\circ \text{ or } 90^\circ$, representing parallel or
perpendicular BAM alignment, respectively. $(\theta, \phi)$ are the standard spherical polar coordinates. Following the argument above,
we only have to sample projection directions in the x-z plane, so $\phi=0^\circ$ and $\theta = 0^\circ, 22.5^\circ, 45^\circ, 67.5^\circ, \text{ or } 90^\circ.$

\subsubsection{PH Patterns}\label{sec:ph-pattern}
Each projection will result in one PH pattern. The goal is to determine whether PH patterns correlate with BAM alignment (see Section~\ref{sec:method}). In the following, we summarize the process of propagating simulation output, namely, data cubes of density ($n$) and B-field ($\vect{B}$), to PHs. The \texttt{YT} (\citealt{2011ApJS..192....9T}) function, \texttt{off\_axis\_projection}, is used to project $n$ along a direction $\vect{\hat{e}}$ into a column density map, $N$, with a 0.02 pc resolution. The resolution of 0.02 pc is comparable to that of a 10-meter submillimeter telescope on a Gould-belt molecular cloud. 

\begin{figure}
    \centering
    \includegraphics[page=5,width=\textwidth]{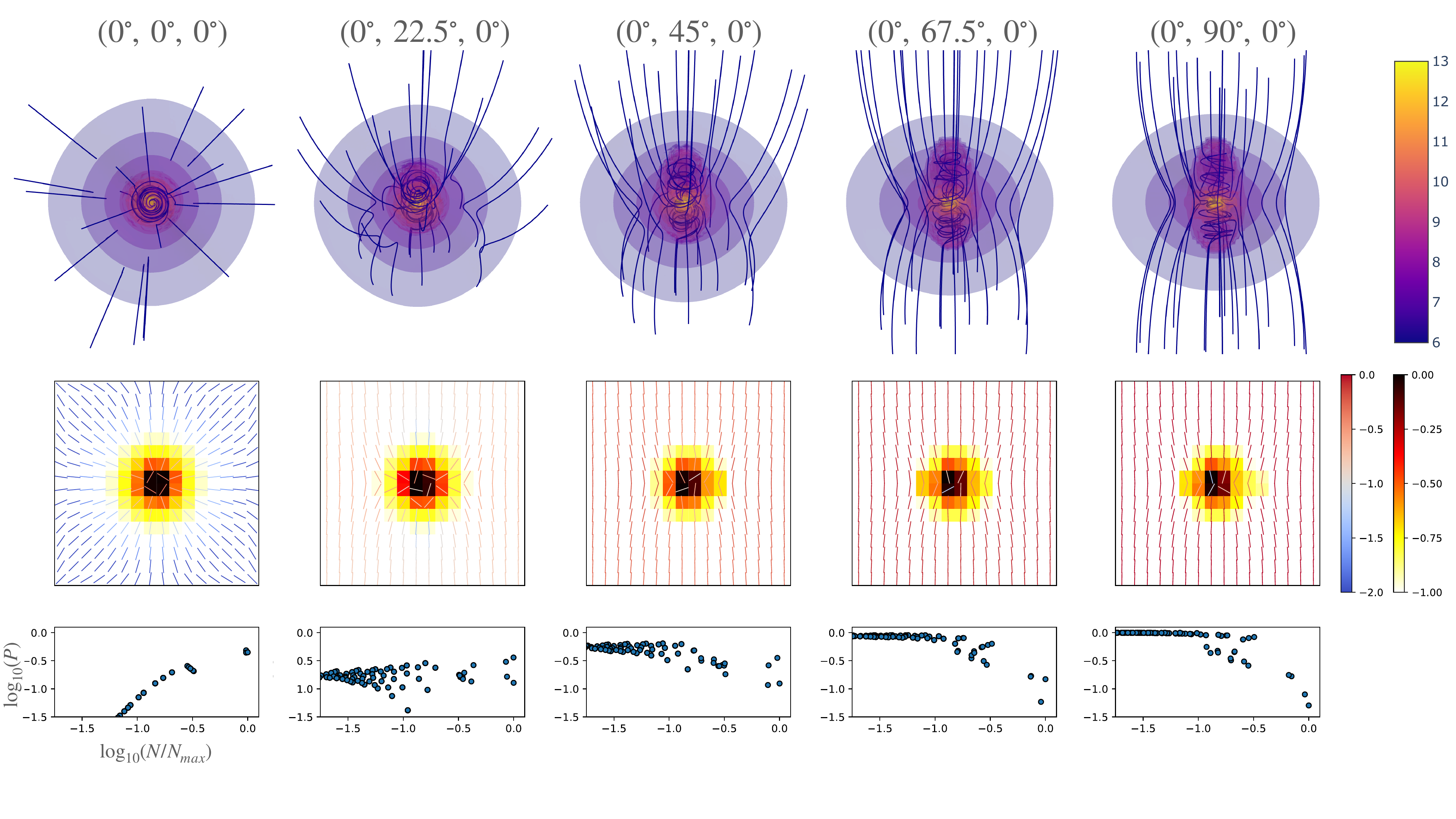}
    \caption{The case of parallel BAM alignment. The snapshot was taken at 0.34 Myr. The projection directions are along ($\theta$, $\phi$) = \{(0$^\circ$, 0$^\circ$), (22.5$^\circ$, 0$^\circ$), (45$^\circ$, 0$^\circ$), (67.5$^\circ$, 0$^\circ$), (90$^\circ$, 0$^\circ$)\}. 
    \textbf{Top}: The 3D views of the core with the lines of sight corresponding to the projection directions. The 3D contours indicate $\log_{10}{n}$ and the lines indicate sampled field lines. 
    \textbf{Middle}: Normalized column density, $\log_{10}(N/N_{\max})$, overlapped with the polarization, $\log_{10}(P)$. The colorbar in
the left/right is for polarization/column density. 
    \textbf{Bottom}:  $\log_{10}(P)$ vs $\log_{10}(N/N_{\max})$. Note that $\theta = 0$ and $22.5^\circ$ do not produce PHs. }
    \label{fig:p1}
\end{figure}

\begin{figure}
    \centering
    \includegraphics[page=6,width=\textwidth]{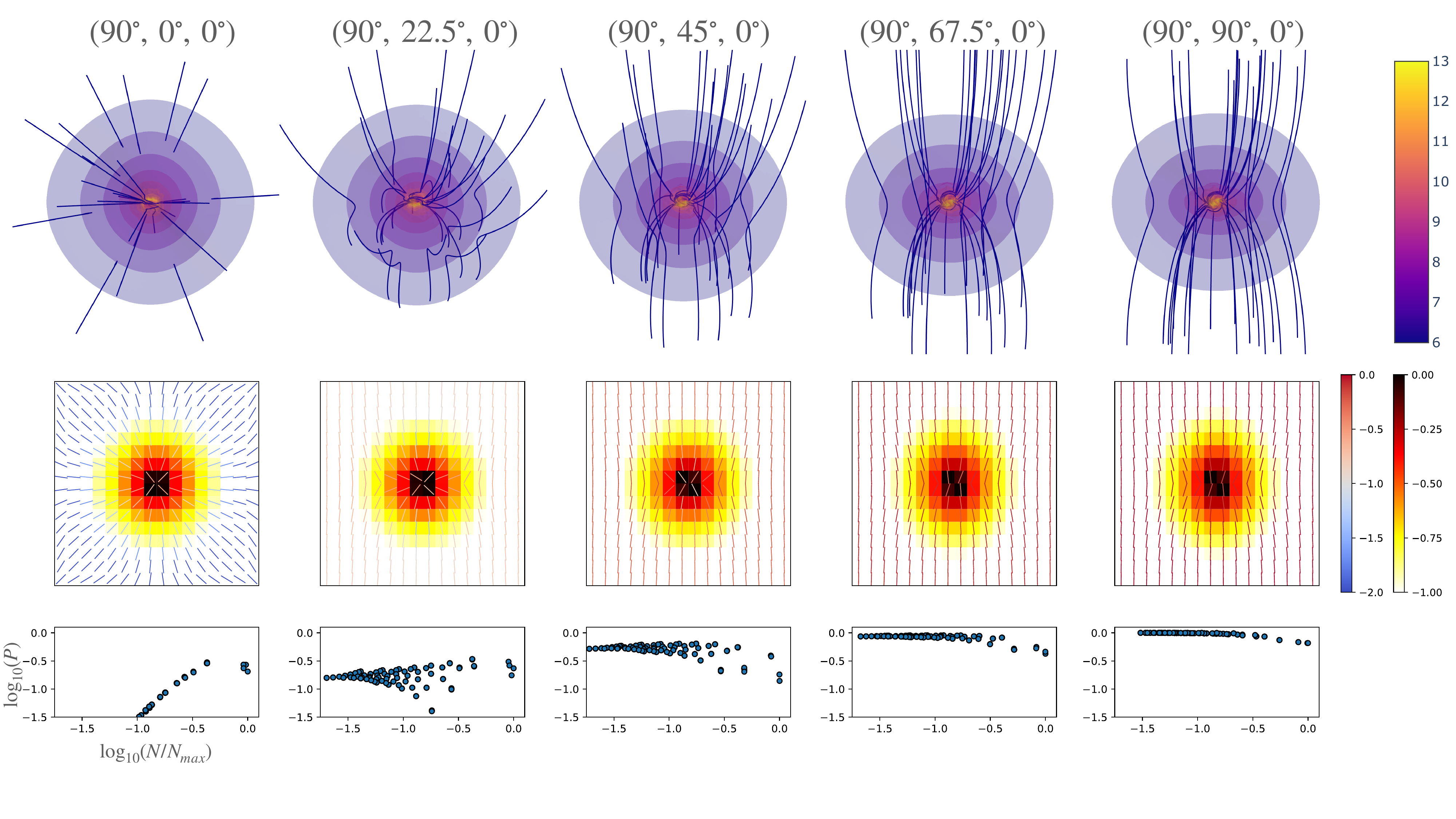}
    \caption{The same as Figure \ref{fig:p1} but for the perpendicular case ($\ell = 90^\circ$) at 0.34 Myr.}
    \label{fig:p2}
\end{figure}

While the B-field-grain alignment mechanism is still not certain, the fact that grains tend to be orientated perpendicular to the local B-field is implied from the fact that dust thermal emission and synchrotron radiation are parallelly polarized \citep{1964BAN....17..465B}. In this case, for ordered B-field, polarization of dust thermal emission should approach zero when LOS ($\vect{\hat{e}}$) approaches the B-field direction, when all directions within the plan of sky (POS) are perpendicular to the B-field with no preference on projected dust grain orientations. So, the polarization fraction should be proportional to $\sin(\hat{\phi})$, where $\hat{\phi}$ is the angle between $\vect{\hat{e}}$ and B-field.
The B-fields within the telescope beam should not always be ordered, of course, which is another major factor of the polarization fraction $P$ of thermal dust emission. More precisely, $P$ depends on the B-field component perpendicular to $\vect{\hat{e}}$, i.e., $\vect{B}_P = \vect{B}-\vect{B} \cdot \vect{\hat{e}}$. The orderliness of $\vect{B}_P$ can be measured by Stokes parameters \citep{chandrasekhar1947transfer} $q = \cos(2\hat{\theta})$ and $u = \sin(2\hat{\theta})$, where $\hat{\theta}$ is the angle direction of $\vect{B}_P$ measured counter-clockwisely from an arbitrary reference direction, $\vect{\hat{e}}'$, within the POS. In the \texttt{YT} calculation above, replacing $n$ with $\textred{\sin^2}(\hat{\phi})\ q$ or $\textred{\sin^2}(\hat{\phi})\ u$ and weighting the value by $n$ yields the $Q$ or $U$ Stokes map. Finally, $P = \sqrt{Q^2 + U^2}$, and the polarization direction measured from $\vect{\hat{e}}'$ is $\frac12\mathrm{atan2}(U, Q)$. Note that this $P$ value is not observable; though we have assumed a constant GAE, it cannot be 100 percents. We should study normalized value, $P/P_{\max}$, in the following analysis to bypass the effect from the unknown GAE; similarly, we will use normalized column density, $N/N_{\max}$.  Figure \ref{fig:p1} shows the overlapped maps of $P$ and $N/N_{\max}$, along with their pixel-by-pixel correlation plots.
\textred{Given that the GAE should not be significantly larger than 10\% \citep{Draine_2009} and that the
sensitivity of a state-of-the-art polarimeter is not better than 0.1\%, $P < 0.01$ is unobservable. Most (95\%) of the $P$'s calculated above are greater than 0.01 (Figure~\ref{fig:p1}). Note that for $\theta < 22.5^\circ$, there will be a polarization ``hump'' instead of hole centered with the column density peak.}


\section{Method \textred{for Classifying Polarization holes}}\label{sec:method}
We first assess the potential classification of the simulated PHs based on their BAM alignment. Only then can the BAM alignments of observed PHs possibly be evaluated by comparing them with the simulated ones.

Cluster analysis, also called data segmentation, aims to segment or group a collection of objects into subsets or clusters, such that those within each cluster are more \emph{closely related} to one another than objects assigned to different clusters. 
For this purpose, we define the dissimilarity between two polarization-hole patterns $\cS_i, \cS_{i'}$ as their Euclidean distance
\begin{equation}
D(\cS_{i}, \cS_{i'}) = \sum_{j=1}^m \left[(\tilde N_{ij}-\tilde N_{i'j})^2 + (\tilde P_{ij}-\tilde P_{i'j})^2\right]\,,
\label{eq:dist}
\end{equation}
\textred{where $\tilde N, \tilde P$ are the normalized column denisty $N$ and the degree of polarization $P$ in the log scale, respectively,
$$
\tilde N_{ij} = \log_{10}\frac{N_{ij}}{\max_{1\le j\le m}N_{ij}}\,,\quad \tilde P_{ij} = \log_{10}\frac{P_{ij}}{\max_{1\le j\le m}P_{ij}}\,;
$$}%
and $\cS_i\triangleq\{(N_{ij}, P_{ij})\}_{j=1}^m$ denotes a polarization-hole pattern characterized by $m$ pixel points. Note that we assume that the pixel points across different patterns have a one-to-one corresponding relationship, so we can define the distance between two patterns as the summation of distances at each pixel point. The subscript $i$ in $\cS_i$ (and $i'$ in $\cS_{i'}$) indexes the projection labels $(\ell,\theta, \phi)$ defined in Section~\ref{sec:sample_proj}. 
We are concerned about whether the parallel cloud ($\ell = 0^\circ$) can be distinguished from the perpendicular cloud ($\ell = 90^\circ$) no matter which $(\theta, \phi)$ is. 

\begin{figure}
    \centering
    \includegraphics[width=\textwidth]{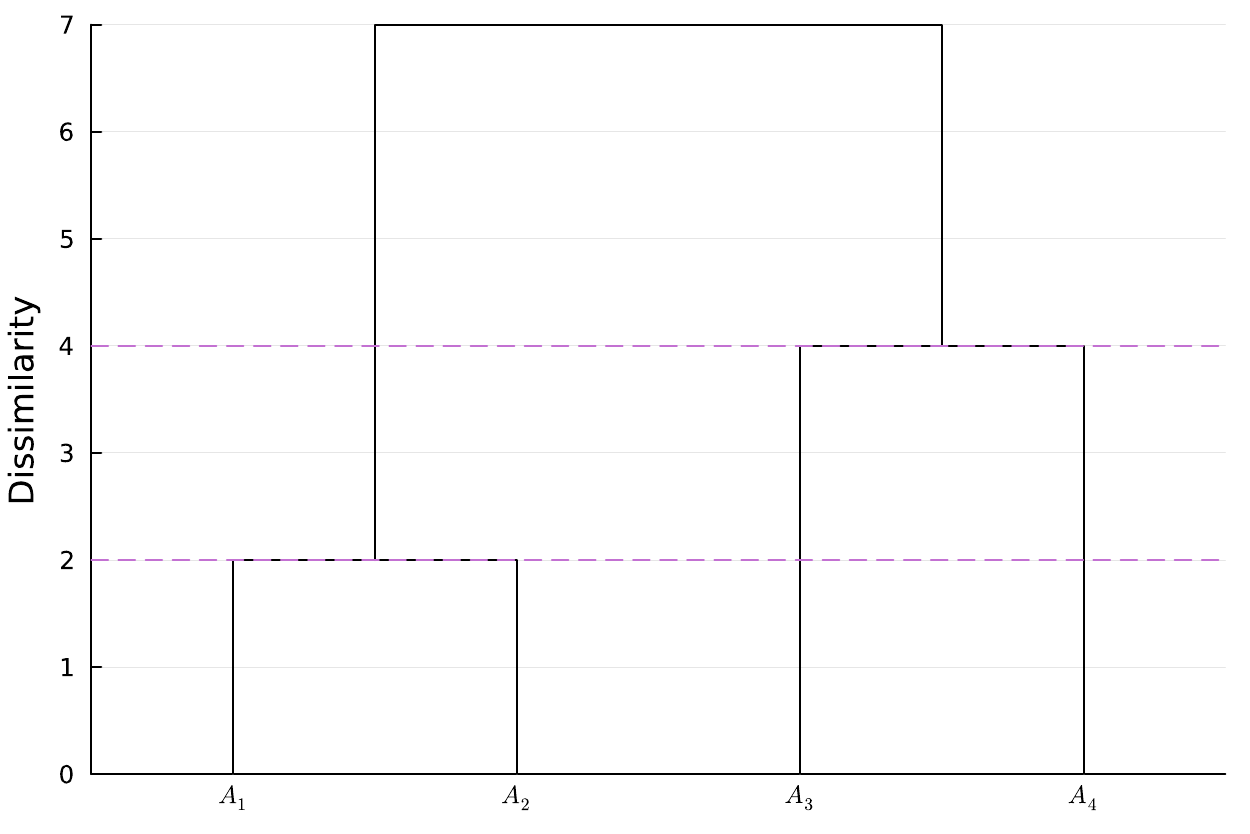}
    \caption{An example of a dendrogram of four objects $(A_1, A_2, A_3, A_4)$. The $y$-axis denotes the dissimilarity $D$ between objects or $D_{\mathrm{SL}}$ between clusters.}
    \label{fig:demo}
\end{figure}

Hierarchical clustering is one of the most popular clustering algorithms. It produces hierarchical representations in which the clusters at each level of the hierarchy are created by merging clusters at the next lower level. An intuitive graphical display is a binary tree, as shown in Figure \ref{fig:demo}, which is also referred to as \emph{dendrogram}. The height of each node is proportional to the value of the intergroup dissimilarity between its two daughters. The terminate nodes representing individual observations are all plotted at zero height. 
For example, at the level (height) of 3, there are 3 clusters, $\{(A_1, A_2), A_3, A_4\}$. If we increase the level to 5, it reduces to 2 clusters since $A_3$ and $A_4$ are merged at the height = 4. At the lowest level (i.e., the zero height), each cluster contains a single object. At the highest level, only one cluster contains all four objects $\{(A_1, A_2, A_3, A_4)\}$. 

Building such a dendrogram usually follows an agglomerative (bottom-up) procedure, which starts at the bottom with each observation (here, a polarization-hole pattern $\cS_i$) representing a singleton cluster. Then at each level recursively merge the closest two clusters into a single cluster, producing one less cluster at the next higher level. Therefore, we need to define a measure of dissimilarity between two clusters. Let $\cG_1$ and $\cG_2$ represent two such clusters. Their dissimilarity $D_{\mathrm{SL}}(\cG_1, \cG_2)$ is defined by the lowest dissimilarity between individual PHs, $D(\cS_i, \cS_{i'})$, where $\cS_i$ is from $\cG_1$ and $\cS_{i'}$ is from $\cG_2$.
$$
D_{\mathrm{SL}}(\cG_1, \cG_2) \triangleq \min_{\cS_i\in \cG_1, \cS_{i'}\in \cG_2} D(\cS_i, \cS_{i'}),
$$
which is also known as the \emph{single linkage} (SL) in the literature \citep{hastieElementsStatisticalLearning2009}.

Once we obtain a dendrogram, if parallel BAM alignments ($\ell = 0^\circ$) can be fully distinguished from perpendicular BAM alignments ($\ell = 90^\circ$), we should be able to cut the dendrogram at some level to form two clusters $\cC_1$ and $\cC_2$,
such that each cluster
only contains a single type of BAM alignment, i.e., all parallel cases are clustered into $\cC_1$ (or $\cC_2$) \emph{and} all perpendicular cases are grouped into $\cC_2$ (or $\cC_1$).
On the other hand, if parallel cases can not be fully distinguished from perpendicular cases, we would try to find the largest PH subset such that parallel and perpendicular BAM alignments can be distinguished within the subset.  The algorithm to find the size of this kind of a subset is given in Appendix~\ref{sec:cluster}. If the subset size is equal to the full sample size, we call the
two types of BAM alignment are fully distinguishable.



\section{Results}\label{sec:res}

\begin{figure}
    \centering
    \includegraphics[width=\textwidth]{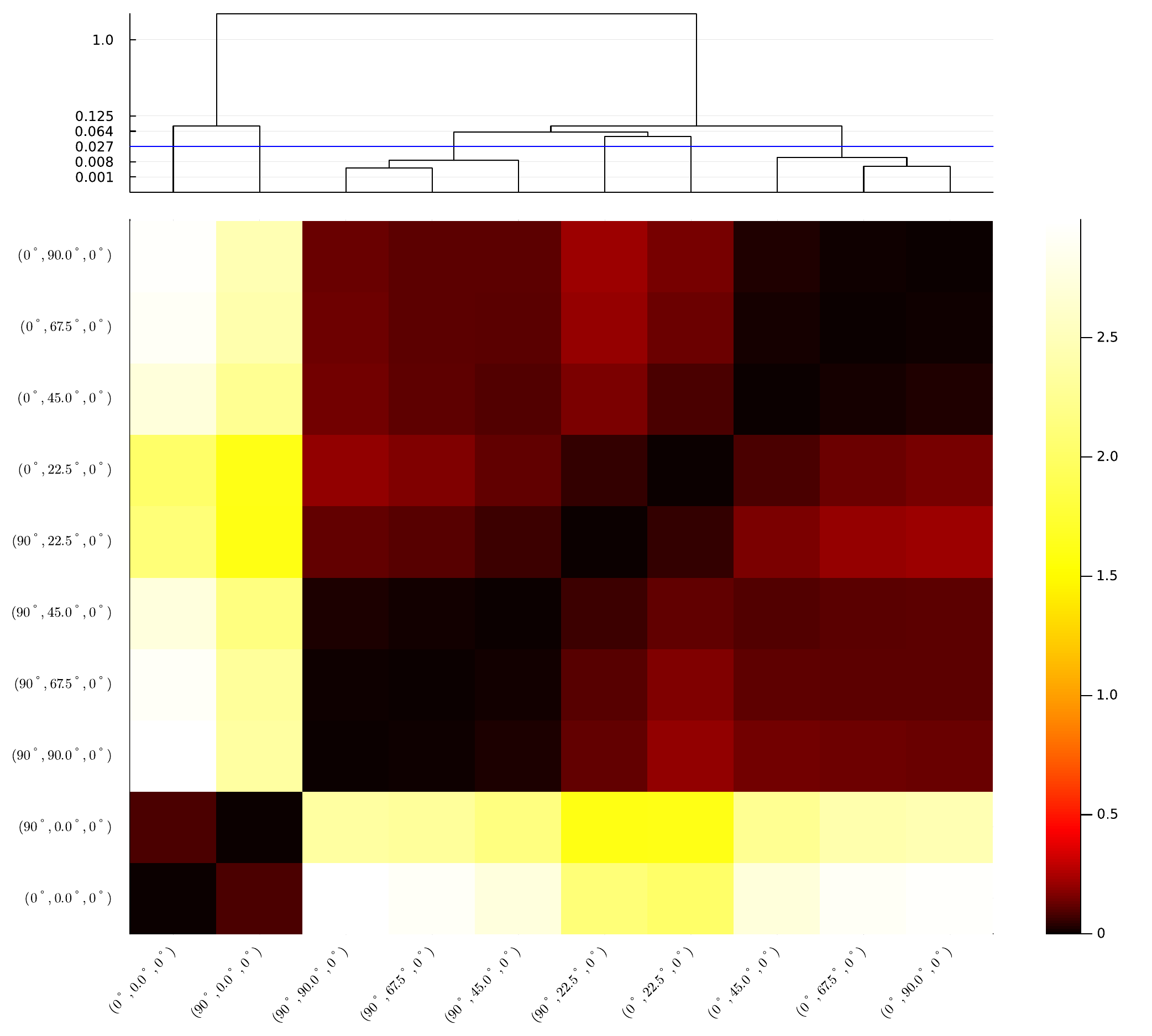}
    \caption{Heatmap of the dissimilarities between all possible pairs of PHs from Figures~\ref{fig:p1} and \ref{fig:p2}, together with the dendrogram for clustering analysis. The axes ticks show the projection labels $(\ell, \theta,\phi)$.}
    \label{fig:snap35}
\end{figure}

\textred{As shown in Figure~\ref{fig:snap35}, there are two non-singleton clusters with $D_{\mathrm{SL}}$ below 0.027, and the two clusters successfully distinguished between different BAM alignments. Singletons excluded by these two clusters have a $\theta$ of $0^\circ$ or $22.5^\circ$, and those with the same $\theta$ cluster at higher $D_{\mathrm{SL}}$. Following the discussion in Section~\ref{sec:ph-pattern} and Figure~\ref{fig:p1} regarding polarization peaks, we can conclude that PHs are indeed distinguishable by $D_{\mathrm{SL}}$.}
Polarization peaks are seldom observed with an apparent reason - the polarization fractions from this perspective are generally lower (see Figures~\ref{fig:p1} and \ref{fig:p2}) \textred{and thus take longer time to detect; usually, the map looks like Figure 1A.} 

Figure~\ref{fig:snap35} is encouraging for us to further explore the idea of using PHs as a method to investigate BAM alignments. In the following, we test this idea by using PHs from snapshots taken 20,000 years earlier to simulate observations.


\begin{figure}
    \centering
    \includegraphics[page=3,width=\textwidth]{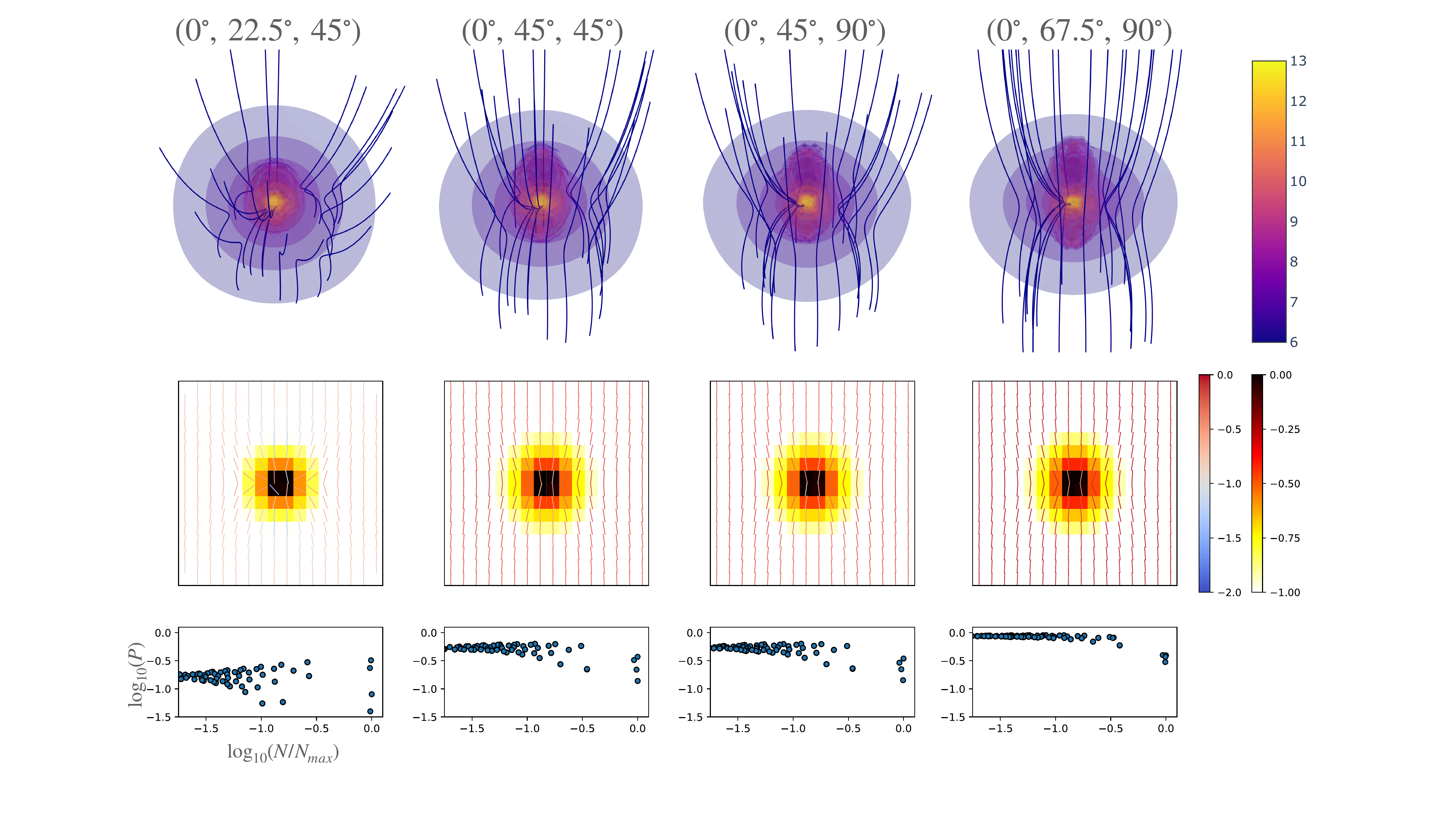}
    \caption{Similar to Figure~\ref{fig:p1}, but taken at 0.32 Myr and projected along ($\theta$, $\phi$) = \{(22.5$^\circ$, 45$^\circ$), (45$^\circ$, 45$^\circ$), (45$^\circ$, 90$^\circ$), (67.5$^\circ$, 90$^\circ$)\}.}
    \label{fig:p3}
\end{figure}

\begin{figure}
    \centering
    \includegraphics[page=4,width=\textwidth]{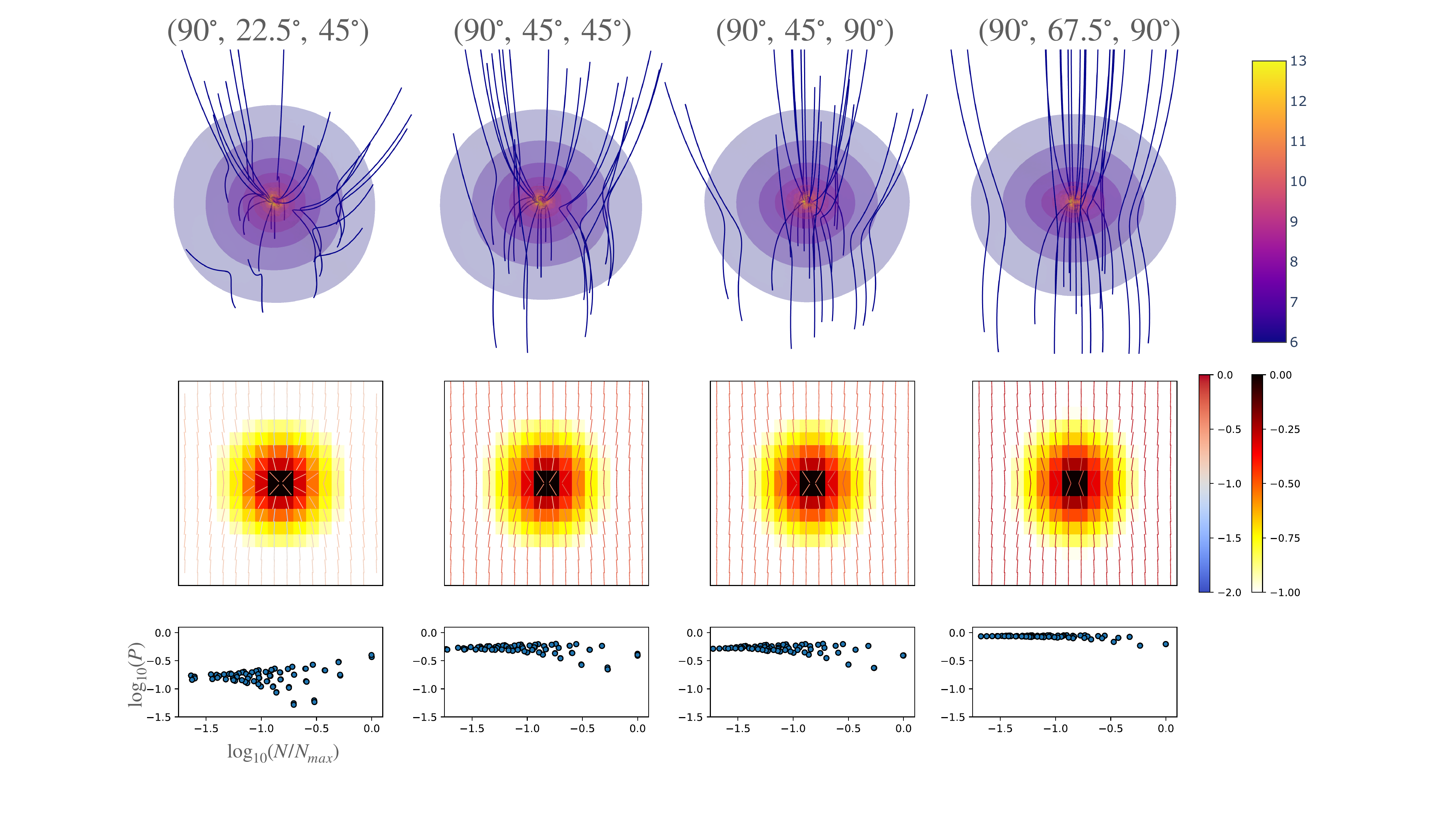}
    \caption{The same as Figure \ref{fig:p3} but for the perpendicular case ($\ell = 90^\circ$).}
    \label{fig:p4}
\end{figure}

We calculate the dissimilarity between the simulated ``observations'' and the benchmark PHs displayed in Figures~\ref{fig:p1} and \ref{fig:p2}, then predict the BAM alignment of each observation based on the alignment of the closest benchmark PH. We consider four projections as displayed in Figures~\ref{fig:p3} and \ref{fig:p4}, which are distinct from those in Figures~\ref{fig:p1} and \ref{fig:p2}, as simulated observations: $(\ell, 22.5^\circ, 45^\circ)$, $(\ell, 45^\circ, 45^\circ)$, $(\ell, 45^\circ, 90^\circ)$, and $(\ell, 67.5^\circ, 90^\circ)$, where $\ell = 0^\circ$ or $90^\circ$. 


In all these cases, the predicted BAM alignments agree with the true alignments, \textred{including the case with $\theta=22.5^\circ$. An encouraging lesson learned here is that even though the dissimilarity between the two benchmark cases with $\theta=22.5^\circ$ is too small to separate them into two clusters (Figure~\ref{fig:snap35}), the smallest distance between an ``observation'' and the benchmark cases can still indicate the BAM alignment even when $\theta=22.5^\circ$.} The details, together with results for simulated observations from further earlier times, can be found in Appendix~\ref{sec:pred}. It should not be difficult to foresee that simulated observations from too early a snapshot will be more difficult to classify, as the cloud rotation has not yet established axial symmetry.





\section{Discussion}\label{sec:discussion}
The previous section shows encouragement. Here we discuss some attention that needs to be paid for the future development of the idea into a practical tool. Our discussion is based on three aspects - simulation, observation and statistics.

\subsection{Simulations}

\textred{
\paragraph{Initial and boundary conditions} In Appendix~\ref{sec:otheryear}, it is evident that the system must evolve for more than 0.3 Myr to become more distinguishable. However, it is important to note that we should only consider the development of symmetry as critical, rather than relying on the surface value of 0.3 Myr. This is because both the initial (Bonnor-Ebert sphere) and boundary (periodic) conditions of the simulations are significantly simplified from reality. In the future, we
shall adopt the results from a simulation of core formation (refer to, e.g., Figure 1 of \cite{Cao_2023}) as a more realistic initial condition. \cite{Cao_2023} accurately replicated the properties of observed molecular clouds, including the ``ordered cloud B fields'' \citep{Li_2009, 2015Natur.520..518L, 2016A&A...586A.138P}, the significantly ``deviated core B fields'' \citep{2014ApJ...792..116Z, 2014ApJS..213...13H, Zhang_2019}, the 2/3 index of the spatial B–n relation \citep{2010ApJ...725..466C, 2020ApJ...890..153J}, magnetic criticality = 1-2 for cores \citep{2013MNRAS.436.3707L, 2015Natur.520..518L, 2021ApJ...917...35M} and core density profiles $n \propto r^{-1.46\pm 0.12}$ \citep{2009ARep...53.1127P, 2013ApJ...765...85K, Zhang_2019}.}

\paragraph{Ambipolar diffusion (AD)} Due to the gas density, B-field strength, and ionization fraction of the disk-formation environment, AD most likely occurs during disk formation. We have also found evidence of turbulence-induced AD (e.g., \citealt{2008ApJ...677.1151L, Tang_2018}). As a first test of the idea of probing BAM alignments with PH patterns, the simulations in this work are ideal MHD, i.e., they do not account for AD. Our MHD code, Scorpio \citep{chengScorpio2023}, is capable of AD simulations, though much more CPU-time-consuming. The success of the ideal-MHD tests encourages us to invest time in future simulations with AD. 

\paragraph{Grain alignment efficiency} Though here we illustrated that PHs do not depend on the decreasing of GAE with increasing density as some literature suggested, GAE may still affect PHs and thus our method. Various GAE theories predict different GAE-density relations. Even though none of them has been tested, we may still apply them to our PH pattern generation process (Section~\ref{sec:sim_res}) to test the robustness of the results in Section~\ref{sec:res}. \textblue{The most reliable piece of information regarding GAE is probably the notable distinction between cloud and disk environments. Recent studies \citep{yangDoesHLTau2019, 2019ApJ...874L...6K} suggest that the Gold-type alignment mechanism, driven by gas flows rather than B-fields \citep{goldAlignmentGalacticDust1952,hoangAlignmentIrregularGrains2018}, appears to be more prevalent in disks. Furthermore, the connection between grain alignments and their sub/millimeter polarization is not well-defined. The increase in complexity of grain size population in disks leads to the polarization not only from thermal emission but also from scattering (e.g., \cite{2018ApJ...865L..12B, 2019MNRAS.482L..29D,stephens2023aligned}). While our simulations focus on disk surroundings and the rotational (disk) component comprises only around 1\% of the total volume, this 1\% may still affect the PH pattern due to its high density. In the future, we will allocate different observed polarization patterns (e.g., aligned with the short projection axis, toroidal, or a combination of the two) to the rotator volume to further assess the robustness of the proposed method.}

\subsection{Observations}

\paragraph{System age} It should be noted that the result in Section~\ref{sec:res} only applies \textred{to systems old enough for developing rotational axial symmetry}. 

\paragraph{Map size} We have also experimented on map sizes. We find that the size must be at least five times as large as the rotator (disk) to cover a substantial portion of the disk envelope, where B-field morphology (and therefore the PH pattern) is more sensitive to BAM alignment. Smaller maps focus more on the rotators, where PH patterns are less distinguishable; see Appendix~\ref{sec:otheryear}.

\subsection{Statistical Methods}

The dissimilarity between two PHs, as defined in Equation~\eqref{eq:dist}, serves only as a proof-of-concept. In reality, however, achieving the one-to-one pixel correspondence required by Equation~\eqref{eq:dist} between observations and simulations is nearly impossible.

With careful observation of the bottom rows of Figures~\ref{fig:p1} and \ref{fig:p2}, one can notice the differences between the $(P, N)$ distributions. First of all, the ``holes'' of the $\ell = 0^\circ$ cases were deeper than those of the $\ell = 90^\circ$ cases. In other words, if we fit the relation between $P$ and $N$ for $\log_{10}(N/N_{\max}) > -1$, the slope will be more negative for the $\ell=0^\circ$ cases. This is because the parallel BAM alignment can entangle the field lines more severely at higher densities, as can be observed by comparing the upper rows of Figures~\ref{fig:p1} and \ref{fig:p2}.

Moreover, fixing $\ell$, we can notice that the $(P,N)$ distributions vary with $\theta$. Cases with $\theta=90^\circ$ and $67.5^\circ$ appear to be the most similar, and the similarity decreases as the difference in $\theta$ increases. These observations agree with the dendrogram in Figure~\ref{fig:snap35}, where we also see that the dissimilarity increases with the $\theta$ difference.

Above discussion inspired us that the $(P, N)$ distributions or ``shapes'' can be the potential signatures of BAM alignments. Most importantly, shape comparison does not require one-to-one pixel correspondence. We have developed a curve-fitting method for this purpose \citep{wangMonotoneCubicBSplines2023b}, and the details will be revealed in a follow-up article. \textred{Once we relax the requirement of pixel-to-pixel correspondence in Equation (3), by convolving MHD simulation results with various beam sizes, we can explore how resolution or distance may impact the PH clustering.}

\textred{%
\subsection{Compared to the Conventional Method for Assessing BAM Alignment}

Traditionally, outflows and B-field directions inferred from polarizations are used to investigate BAM alignment, assuming that outflows are aligned with AM (e.g., \citealt{Hull_2013,Yen_2021}). There are several shortcomings of this method that the proposed PH method does not share.

First, the traditional method can only be applied to cases with outflows, which may lead to bias. For example, when comparing the top rows of Figures \ref{fig:p1} and \ref{fig:p2}, we can easily observe bipolar outflows in Figure \ref{fig:p1} but not in Figure \ref{fig:p2}. The method is only valid when outflows do not favor specific BAM alignment.

Second, the traditional method depends on the shape of the cumulative distribution function (CDF) of outflow-field angles projected on the sky, assuming a single-peaked distribution of the 3D BAM alignment. However, in reality, the 3D alignment might be bimodal or even multi-modal; the proposed PH clustering method can resolve these conditions. In the upcoming sequel, we will incorporate multi-modal conditions instead of just bimodal (0 and 90 degrees).

Last but not least, the traditional method strictly relies on the CDF of a large sample, and no conclusions can be drawn for individual cases. On the other hand, single PH can still be compared to the benchmark simulations for studying BAM alignment.    
}

\section{Conclusion}\label{sec:conclusion}
With MHD numerical simulations, we intend to study the difference in B-field structures stemming from different BAM alignments. We use the PHs resulting from the projections of B-fields as the probe, which is a brand-new approach. The result is encouraging -- PHs can be classified based on their BAM alignments once the rotation system is mature enough to exhibit rotational symmetry. This means that PHs can potentially serve as probes for the alignment between disks and B-fields. This alignment has been suggested as a solution to the magnetic braking catastrophe during disk formation. The next step is to incorporate additional factors, such as variations in observational resolution and ambipolar diffusion, into the test in order to further develop the PH idea into a practical probe of BAM alignments.

\section*{Acknowledgment}
This research is supported by General Research Fund grants from the Research Grants Council of Hong Kong (14303819, 14307222 and 14303921), and by Collaborative Research Fund grant (C4012-20E). Lĳun Wang was
supported by the Hong Kong Ph.D. Fellowship Scheme from the University Grant Committee.

%






\appendix

\section{Determining the Largest PH Subset that is fully distinguishable in BAM alignment}\label{sec:cluster}

If $\ell = 0^\circ$ can be fully distinguished from $\ell=90^\circ$, then the resulting clustering label $\hat\ell$, taking 1 or 2, should be consistent with the offset labels, i.e., $\hat\ell=1$ corresponds to $\ell=0^\circ$ (or $\ell = 90^\circ$). We use $I(\cdot)$ to denote the indicator function.
Since the cluster structure is invariant to the permutation of the label symbols, we define the accuracy as
$$
\Acc(\ell, \hat\ell) \triangleq \max_{\phi\in\Phi} \frac 1n\sum_{i=1}^n I(\phi(\hat\ell_i) = \ell_i)\,,
$$
where $n$ is the number of observations and $\Phi = \{\phi: \phi \text{ is a bijection from $\{1, 2\}$ to $\{0^\circ, 90^\circ\}$}\}$.

If parallel clouds ($\ell = 0^\circ$) can not be fully distinguished from perpendicular clouds ($\ell=90^\circ$) among all projections, we would try to find the largest projection set such that parallel clouds and perpendicular clouds can be distinguished. Let the full projection set be $\cP_0$. Let the size difference between the full projection set and a candidate projection set $d=\vert \cP_0\vert -\vert \cP\vert$. Then, the goal to determine the largest set can be written as
\begin{equation}
\min_d \max_{\cP\subseteq \cP_0, \vert \cP\vert=\vert\cP_0\vert-d} \Acc(\ell_\cP, \hat\ell(\cP))\,,    
\label{eq:minmax}
\end{equation}
where $\ell_\cP$ denotes the subvector of $\ell$ for the projection set $\cP$. If the solution to Equation~\eqref{eq:minmax} is $\hat d = 0$, then $\ell = 0^\circ$ is fully distinguishable from $\ell = 90^\circ$ among all projections. Algorithm~\ref{alg:sol} summarizes the procedure to solve Equation~\eqref{eq:minmax}. 

\begin{algorithm}[H]
\caption{Largest Projection Set for Distinguishable Clouds}\label{alg:sol}
\begin{algorithmic}[1]
    \FOR{$d=0,1,2,\ldots,\vert \cP_0\vert-1$}
    \FOR{each subset $\cP$ with $\vert \cP_0\vert-d$ elements}
    \STATE Perform clustering and measure the accuracy $\Acc(\ell_\cP,\hat\ell(\cP))$
    \ENDFOR
    \STATE Pick $\cP^\star$ which achieves the highest accuracy
    \IF{Acc == 1.0}
    \STATE break
    \ENDIF
    \ENDFOR
\end{algorithmic}
\end{algorithm}

\section{BAM alignment Prediction for simulated Observations}\label{sec:pred}

Compared to the benchmark PHs (Figures~\ref{fig:p1} and \ref{fig:p2}, taken at 0.34 Myr), the simulated observations were taken from earlier snapshots of the simulations and projected differently at four directions: $(\ell, 22.5^\circ, 45^\circ)$, $(\ell, 45^\circ, 45^\circ)$, $(\ell, 45^\circ, 90^\circ)$, and $(\ell, 67.5^\circ, 90^\circ)$). Note that the benchmark PHs are with LOS within the x-z plane, assuming the B-field geometry is rotational axial symmetric after the system stabilized. For simulated observations, we chose LOS outside the x-z plane ($\phi>0^\circ$).  We predict the BAM alignment of an observation by that of its closest benchmark PHs. Tables~\ref{tab:pred34}, ~\ref{tab:pred33} and~\ref{tab:pred32} show the closest benchmark PH for each simulated observation from snapshots at  
0.33 Myr, 
0.32 Myr, 
and 0.31 Myr
, respectively.

\begin{table}
    \centering
    \caption{The closest benchmark PHs for observations taken at 0.33 Myr.}
    \begin{tabular}{ccccc}
\toprule
&$(\ell, 22.5^\circ, 45^\circ)$&$(\ell, 45^\circ, 45^\circ)$&$(\ell, 45^\circ, 90^\circ)$&$(\ell, 67.5^\circ, 90^\circ)$\tabularnewline
\midrule
$\ell=0^\circ$&$(0^\circ, 22.5^\circ, 0^\circ)$&$(0^\circ, 45^\circ, 0^\circ)$&$(0^\circ, 45^\circ, 0^\circ)$&$(0^\circ, 67.5^\circ, 0^\circ)$\tabularnewline
$\ell=90^\circ$&$(90^\circ, 22.5^\circ, 0^\circ)$&$(90^\circ, 45^\circ, 0^\circ)$&$(90^\circ, 45^\circ, 0^\circ)$&$(90^\circ, 67.5^\circ, 0^\circ)$\tabularnewline
\bottomrule
\end{tabular}

    \label{tab:pred34}
\end{table}

\begin{table}[H]
    \centering
    \caption{The closest benchmark PHs for observations taken at 0.32 Myr.}
    \begin{tabular}{ccccc}
\toprule
&$(\ell, 22.5^\circ, 45^\circ)$&$(\ell, 45^\circ, 45^\circ)$&$(\ell, 45^\circ, 90^\circ)$&$(\ell, 67.5^\circ, 90^\circ)$\tabularnewline
\midrule
$\ell=0^\circ$&$(0^\circ, 22.5^\circ, 0^\circ)$&$(0^\circ, 45^\circ, 0^\circ)$&$(0^\circ, 45^\circ, 0^\circ)$&$(0^\circ, 67.5^\circ, 0^\circ)$\tabularnewline
$\ell=90^\circ$&$(90^\circ, 22.5^\circ, 0^\circ)$&$(90^\circ, 45^\circ, 0^\circ)$&$(90^\circ, 45^\circ, 0^\circ)$&$(90^\circ, 67.5^\circ, 0^\circ)$\tabularnewline
\bottomrule
\end{tabular}

    \label{tab:pred33}
\end{table}

\begin{table}[H]
    \centering
    \caption{The closest benchmark PHs for observations taken at 0.31 Myr.}
    \begin{tabular}{ccccc}
\toprule
&$(\ell, 22.5^\circ, 45^\circ)$&$(\ell, 45^\circ, 45^\circ)$&$(\ell, 45^\circ, 90^\circ)$&$(\ell, 67.5^\circ, 90^\circ)$\tabularnewline
\midrule
$\ell=0^\circ$&$(90^\circ, 22.5^\circ, 0^\circ)$&$(0^\circ, 45^\circ, 0^\circ)$&$(0^\circ, 45^\circ, 0^\circ)$&$(0^\circ, 67.5^\circ, 0^\circ)$\tabularnewline
$\ell=90^\circ$&$(90^\circ, 22.5^\circ, 0^\circ)$&$(90^\circ, 45^\circ, 0^\circ)$&$(90^\circ, 45^\circ, 0^\circ)$&$(90^\circ, 67.5^\circ, 0^\circ)$\tabularnewline
\bottomrule
\end{tabular}

    \label{tab:pred32}
\end{table}

All predictions are accurate for observations at 0.33 Myr and 0.32 Myr, but it becomes less accurate for the snapshot at 0.31 Myr, where the observation with $(0^\circ, 22.5^\circ, 45^\circ)$ is misclassified with the benchmark PH with $(90^\circ, 22.5^\circ, 0^\circ)$.

\section{Explorations of Rotational Radius and Age}\label{sec:otheryear}

As mentioned in Section~\ref{sec:mhdsim}, the $r_{rot}$ is set at 0.03 pc, which corresponds to 20\% of the core size ($r_{core}$). We emphasized that the value of $r_{core}$ needs to be significantly larger than $r_{rot}$ in order to distinguish the PHs with different BAM alignments. It is crucial to observe PH patterns outside the rotational part, i.e., the envelope of a disk, not the disk itself. To illustrate this, we define
$$
\ratio = \frac{r_{rot}}{r_{core}}\,,
$$
and demonstrate that a ratio as low as 0.2 is required to achieve greater accuracy in distinguishing BAM alignments.

Also, as mentioned above, the rotation needs to be sufficiently developed for the BAM alignments to be distinguishable. A rotation that is too young cannot fully develop axial symmetry, making it more difficult to classify the PH patterns. We demonstrate this with different years of the simulation.


For each \ratio and year, we checked whether parallel BAM alignments ($\ell = 0^\circ$) can be distinguished from perpendicular alignments ($\ell=90^\circ$) among all PHs. Equivalently, we would find the largest PH subset $\cP^\star$ that is fully distinguishable  (i.e., $\Acc(\ell,\hat\ell)=1.0$). 


\begin{figure}
    \centering
    \includegraphics[width=\textwidth]{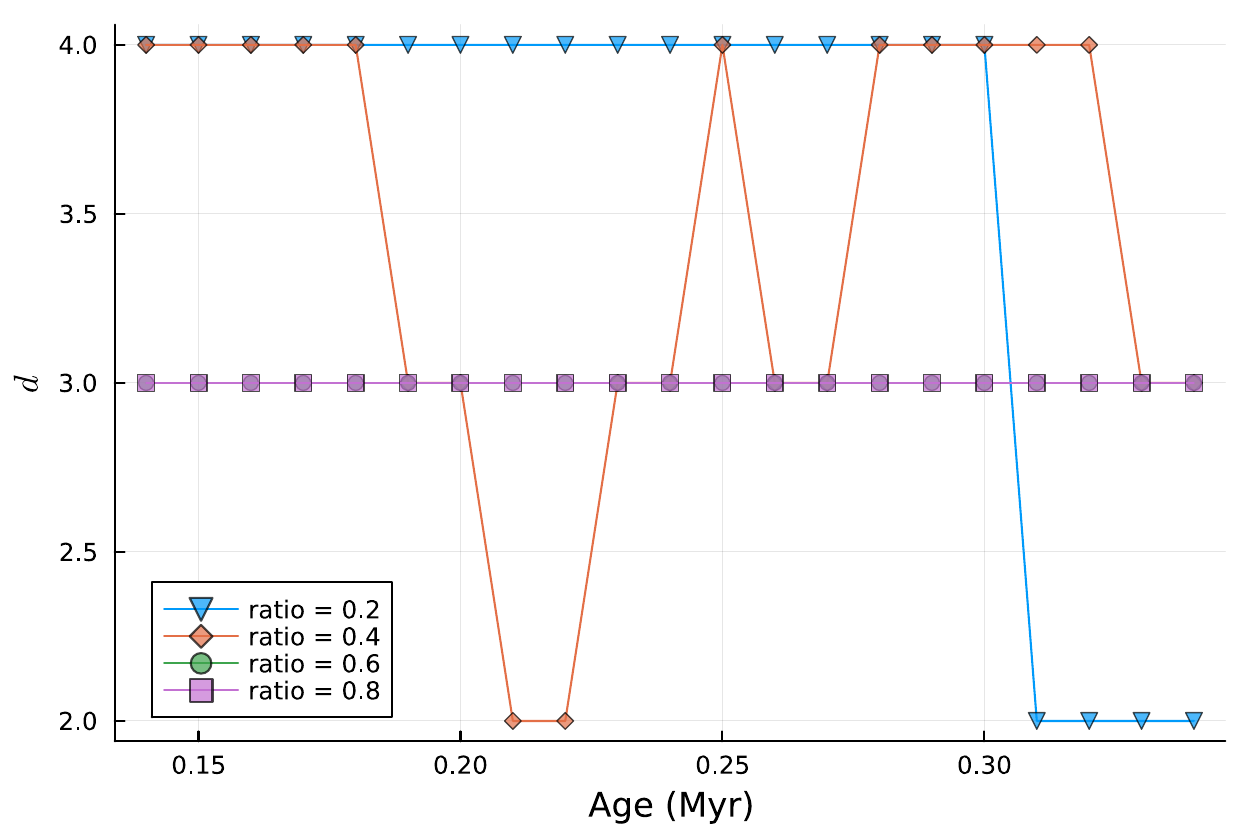}
    \caption{The vertical axis shows the size difference $d$ 
 between the full PH set and the largest subset for achieving the highest accuracy (1.0).}
    \label{fig:d_acc}
\end{figure}

Following Algorithm~\ref{alg:sol}, the resulting size difference $d = \vert \cP_0\vert - \vert\cP^\star\vert$ is summarized in Figure~\ref{fig:d_acc}. Note that the minimum $d$ is 2, due to that the cases with $\theta = 0^\circ$ and $22.5^\circ$ do not possess a PH (see Figures~\ref{fig:p1} and \ref{fig:p2} and Section~\ref{sec:res}). 

\begin{figure}
    \centering
    \includegraphics[width=\textwidth]{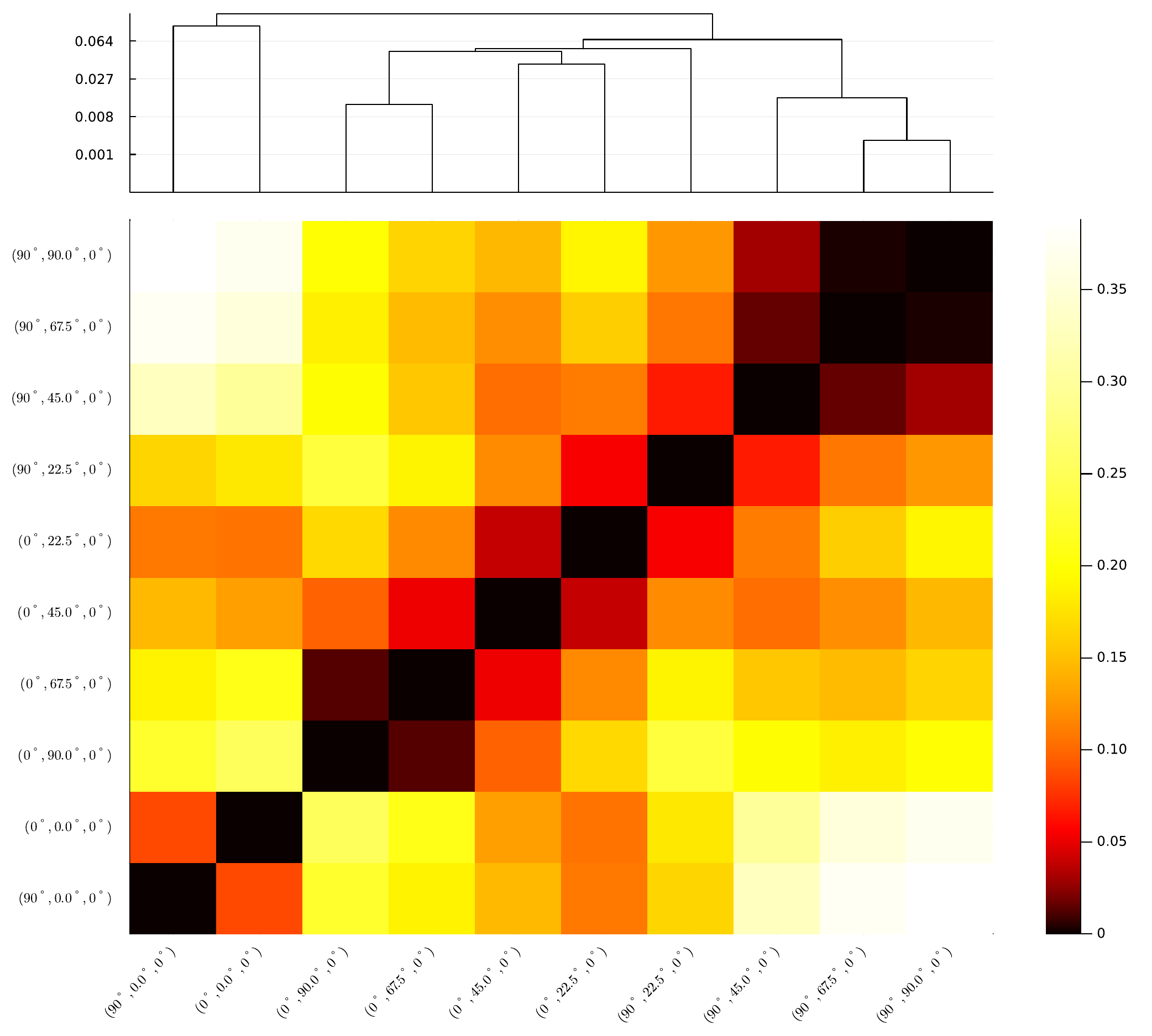}
    \caption{The same plot to Figure \ref{fig:snap35} but with $\log_{10}(N/N_{\max})>-1.0$.}
    \label{fig:heatmp-1.0}
\end{figure}

\textred{Another approach to understanding the effect of map size is to consider the ratio = 0.2 case and cut the map to a lower limit, $N_{\min}$. Cutting at $N_{\min}/N_{\max} = 10^{-1.5}$ results in minimal alteration to Figure~\ref{fig:snap35}, whereas cutting at $10^{-1}$ significantly impacts the distinguishability, as depicted in Figure~\ref{fig:heatmp-1.0}.}

\bibliography{sample631, Astro_citations, Pairwise}{}
\bibliographystyle{aasjournal}



\end{document}